\documentclass{INTERSPEECH2023}

\interspeechcameraready

\title{Sequence-to-Sequence Multi-Modal Speech In-Painting}
\name{Mahsa Kadkhodaei Elyaderani$^1$, Shahram Shirani$^2$}
\address{
  $^1$Department of Computational Science and Engineering, McMaster University, Canada\\
  $^2$Department of Computational Science and Engineering, McMaster University, Canada}
\email{kadkhodm@mcmaster.ca, shirani@mcmaster.ca}

\begin{document}

\maketitle
 
\begin{abstract}
Speech in-painting is the task of regenerating missing audio contents using reliable context information. Despite various recent studies in multi-modal perception of audio in-painting, there is still a need for an effective infusion of visual and auditory information in speech in-painting. In this paper, we introduce a novel sequence-to-sequence model that leverages the visual information to in-paint audio signals via an encoder-decoder architecture. The encoder plays the role of a lip-reader for facial recordings and the decoder takes both encoder outputs as well as the distorted audio spectrograms to restore the original speech. Our model outperforms an audio-only speech in-painting model and has comparable results with a recent multi-modal speech in-painter in terms of speech quality and intelligibility metrics for distortions of 300 ms to 1500 ms duration, which proves the effectiveness of the introduced multi-modality in speech in-painting.
\end{abstract}
\noindent\textbf{Index Terms}: speech enhancement, speech in-painting, sequence-to-sequence models, multi-modality

\section{Introduction}

\label{sec:intro}
Auditory information loss may be caused by various sources such as transmission problems, e.g. voice-over-IP connection issues, or noise contamination problems, e.g. corrupted physical devices. Even though enhancing audio signals is an important processing task, it has remained a challenge when corruptions are of long durations ($> 200$ ms). An audio in-painter generates missing parts of audio from the available context. The first attempts in audio in-painting used the help of auto-regressive models, which estimate the missing samples as linear combinations of their neighbours \cite{3}, \cite{4}. The self-similarity approach in \cite{self} exploited available user records to fill in missing audio parts. The recurring musical structures were leveraged in \cite{graph} to learn internal redundancies that exist in time-frequency representations and recover long-duration distortions in musical pieces. Sparse representation modelling in audio in-painting was employed in \cite{1} and \cite{2} to approximate each audio frame as a sparse linear combination of the columns of a dictionary.

Among neural network-based methods, recently \cite{Kagler} proposed a convolutional U-net for suppressing noise injected into temporal or spectral dimensions of speech spectrograms. The approach of \cite{Kagler} employed a VGG-like deep feature extractor, called SpeechVGG, which was obtained by pre-training the famous VGG model for classifying the 1000 most-frequently spoken words in their training dataset. Since spectrograms can be viewed as two-dimensional images, the method of \cite{Chang} tackled the task of filling in missing contents of an audio signal by applying an image in-painting technique, like in \cite{Yu}, to the audio spectral representation. The major innovation of \cite{Chang} was that they replaced standard convolution layers with either gated convolutions (in the case of audio waveform in-painting) or dilated/strided convolutions (in the case of spectrogram in-painting). Similar to \cite{Kagler}, \cite{Chang} investigated the use of a perceptual loss function.

The authors of \cite{masking} proposed two models for uninformed audio in-painting that consisted of down-sampling, residual, and up-sampling blocks to learn and in-paint the locations of noise in the audio. They applied partial convolutions to make the convolution of masked spectrograms only dependent on uncorrupted pixels. Incorporating different modalities into audio in-painting provided complementary information about the signal and enabled robust inference in \cite{multimodalcombine}. Multi-modality was also shown to be successful in the audio-visual correspondence learning method of \cite{later}. The proposed method of \cite{Morrone} incorporated video features into the audio in-painting task to recover speech gaps ranging from 100 ms to 1600 ms using Long Short-Term Memory (LSTM) networks.

Lately, Generative Adversarial Networks (GANs) were applied for in-painting missing audio content. Authors of \cite{gacela} introduced a conditional GAN that used contextual information, along with multiple discriminators of different receptive fields, to distinguish real spectrograms from fake ones. In \cite{wgan}, a Wasserstein GAN was proposed to restore missing audio from adjacent intact regions by minimizing the Wasserstein distance between the ground-truth and the generated data distributions. The multi-modal GAN architecture proposed by \cite{multimodalinfuse} for audio in-painting consisted of a convolutional encoder-decoder model that worked in the joint audio-visual feature space to reconstruct missing audio disruptions of up to 800 ms. The method from \cite{multimodalinfuse} leveraged the WaveNet generative model \cite{wavenet} to decode spectrogram outputs into high-quality audio waveforms. Despite the reported success of GAN-based models for audio in-painting, the training process of these models was computationally challenging.

While many state-of-the-art audio in-painting methods perform over music or environmental sound signals, in this paper, we propose a novel method for recovering speech signals. Inspired by the models in image in-painting and machine translation \cite{seq2seq}, our contribution is to "translate" the visual modality into the auditory modality and reconstruct the missing or corrupted pieces when the locations of such distortions are known. Our sequence-to-sequence model consists of an encoder, to incorporate lip motion features from the videos, and a decoder, for audio spectrograms. Both encoder and decoder are made of stacked bi-directional LSTM (BLTSM) layers. The remainder of the paper is organized as follows. In Section \ref{sec:method}, we elaborate on our sequence-to-sequence multi-modal approach. In Section \ref{sec:res}, we discuss our ablation studies and compare our method with another recent multi-modal speech in-painter, and finally, in Section \ref{sec:con} our paper's contributions are concluded and potential plans to improve our current method are suggested.

\section{Proposed Method}
\label{sec:method}

\begin{figure}[t]
  \centering
  \includegraphics[width=\linewidth]{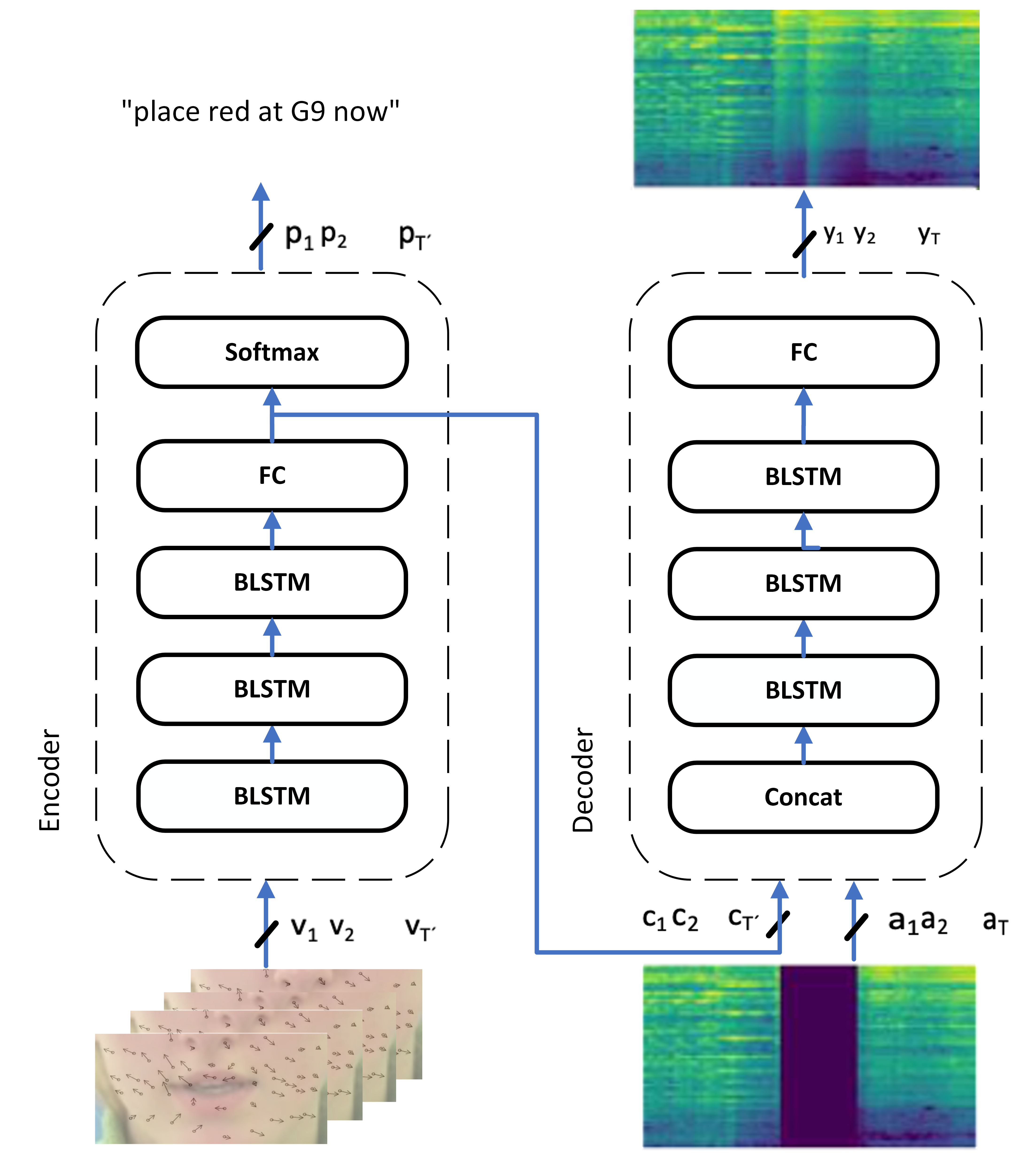}
  \caption{An illustration of the proposed sequence-to-sequence model for speech in-painting. The encoder (in the left) takes the motion vectors from cropped video frames and the decoder (in the right) takes both spectrograms and visual features from the encoder. The model outputs in-painted spectrograms plus corresponding transcriptions.}
  \label{fig:diag}
\end{figure}

In designing a speech in-painting model, it is important to consider the sequential nature of speech signals. The sequence-to-sequence model introduced in this paper has an encoder-decoder architecture to recover missing short words or corrupted phonemes of speech signals by using the corresponding intact videos. Assume $X=[ x_1, x_2, \cdots , x_T ]$ is the spectrogram of a clean speech signal, where $x_t$ for $1\le t \le T$ denotes the frequency vector corresponding to time $t$. The corrupted frequency vector $a_t$ is obtained by multiplying $x_t$ by the mask $m_t \in \{ 0, 1 \} $, i.e. $a_t = m_t \cdot x_t$. Moreover, let us assume $Y=[y_1, y_2, \cdots , y_T]$ denotes the outputs of the decoder. The frequency vectors of the signal spectrogram can then be reconstructed as $$o_t = m_t \cdot x_t + (1 - m_t) \cdot y_t,$$ where $1\le t \le T$. We assume the locations of the masked parts of the spectrogram are known to our method, making it an informed in-painting method. Figure \ref{fig:diag} illustrates the overview of our proposed method, which will be explained in details through the rest of this section.

\subsection{Encoder}

The strong relationship between phonemes of a language and their visual articulations, i.e. shapes of the mouth, forms the main idea behind multi-modal speech in-painting. However, the challenge is that multiple phonemes can have similar mouth shapes, despite their distinguished sounds. As a result, a robust lip reading module exploits the contextual data to differentiate visually-identical phonemes. In our sequence-to-sequence model, the encoder acts as a lip reading module inspired by the LipNet model in \cite{lipnet} and the Lipreading model in \cite{wlipreading}. Originally, both of \cite{lipnet} and \cite{wlipreading} methods were designed to transcribe videos from speakers' lips movements via minimizing the Connectionist Temporal Classification (CTC) loss \cite{lipnet} over the Grid Corpus \cite{grid}. More precisely, the method from \cite{lipnet} exploited spatiotemporal convolutional and BLSTM layers on video frames, while \cite{wlipreading} approach used LSTMs only.

As depicted in Figure \ref{fig:diag}, the output of our sequence-to-sequence model, $y_1, y_2, \cdots , y_T$, is a function of both visual features $v_1, v_2, \cdots v_{T^{\prime}}$ and auditory cues $ a_{1}, a_{2}, \cdots a_{T}$. In our sequence-to-sequence model, the role of the encoder is to transform visual features into a sequence of vectors that are helpful for the decoder to regenerate audio spectrograms at each time-step. As a result, the inputs to the encoder are obtained from cropped mouth-centered video frames. Similar to \cite{morroneface}, facial landmarks are extracted from all cropped video frames and their differences across consecutive frames are fed as landmark motion vectors to the encoder.
Then, BLSTM layers are applied to the computed motion vectors in order to leverage the temporal dependency of extracted visual features. The last BLSTM layer outputs a sequence of hidden states for each input time step. A fully-connected (FC) layer is responsible for outputting features that help the decoder generates missing audio segments, while the top softmax encoder layer generates phoneme probabilities that will be used in computing the CTC loss.

\subsection{Decoder}

The decoder part of our sequence-to-sequence model in-paints spectrogram representations of degraded audios. The input to the decoder is the concatenation of the spectral features and the outputs of the last fully-connected layer of the encoder. The decoder estimates the conditional probability of reconstructed spectrogram vectors, given the input visual and audio features:
\begin{equation}
p(y_{1}, y_{2}, \cdots y_{T} | a_{1}, a_{2}, \cdots a_{T} , v_1, v_2, \cdots v_{T^{\prime}}).
\label{eq:lstm1}
\end{equation}

As presented in Equation~\ref{eq:lstm1}, the audio sequence, $ a_{1}, a_{2}, \cdots a_{T} $, and the video sequence, $v_1, v_2, \cdots v_{T^{\prime}}$, can be of different lengths $T$ and $T^{\prime}$, respectively. In our setup, $T > T^{\prime}$, originally. Therefore, before feeding the video frames to the encoder, we up-sample the frames so that their number matches the temporal length of input spectrograms, i.e., we make sure $T=T^{\prime}$, after up-sampling.
Assuming the outputs of the last fully-connected layer of the encoder are denoted by $c_1, c_2, \cdots, c_{T^{\prime}}$, then our model estimates the above probability by the following expression 
\begin{equation}
p(y_{1}, y_{2}, \cdots y_{T} | a_{1}, a_{2}, \cdots a_{T} , c_{1}, c_{2}, \cdots, c_{T^{\prime}}).
\label{eq:lstm2}
\end{equation}

The encoder outputs $c_{1}, c_{2}, \cdots, c_{T^{\prime}}$ and spectral vectors $a_{1}, a_{2}, \cdots a_{T}$ are temporally concatenated and fed into three BLSTM layers and a FC layer.

The encoder and decoder modules are trained in an end-to-end fashion to minimize a joint loss function that allows the model to learn useful visual and auditory features for speech in-painting. Specifically, we minimize a weighted sum of the mean-squared error (MSE) between the reconstructed and ground-truth spectrograms and the CTC loss for predicting the sequence of spoken phonemes:
\begin{equation}\label{eq:loss}
\text{loss}=\text{MSE} +\lambda \cdot \text{CTC},
\end{equation}
where $\lambda$ is a trade-off parameter.

\section{Experimental Results}
\label{sec:res}

\subsection{Dataset}

The Grid Corpus \cite{grid} is a large multi-speaker audio-visual sentence corpus. Numerous speech perception studies employ the Grid Corpus in their research. There are 34 speakers (18 male, 16 female) in this dataset,  speaking around 1000 sentences. Along with the audio, facial recordings, and aligned transcripts are stored in this dataset as well. Since the speaker 21 in the Grid Corpus has incomplete data, we put the speaker 21 aside and use data of 33 speakers in this paper. The length of each video sample is only 3 seconds. Audio and video rates are 25kHz and 25fps, respectively. Sentences in this corpus follow a pre-determined structure made of six-word categories in the order of command + color + preposition + letter + digit + adverb. Each category has a few word choices. The command can be selected from four \{bin, lay, place, set\} verbs. The color comes from the set \{blue, green, red, white\}. The preposition and adverb are one of the options from \{at, by, in, with\} and \{again, now, please, soon\}. Letters and digits are from \{A, ..., Z\} $\setminus$ \{W\} and \{zero, ..., nine\}, respectively. For instance, one spoken sentence could be ''place red at G9 now". 

\subsection{Data Preparation}

To prepare the data for our experiments, we pre-process both audio and video modalities. First, the audio signals are re-sampled from 25kHz to 8kHz and a pre-emphasis filter is applied to signals to boost high frequencies. Then, the audio Mel-spectrograms are computed by performing the Short-Term Fourier Transform (STFT) with 320 sample points (40 ms) and a hop size of 160 points (20 ms). The length of the windowed signal after padding with zeros is 510 samples corresponding to a duration of 63.75 ms. The magnitude of the STFT is then transformed to Mel scale using 64 Mel filter banks that is followed by log dynamic range compression. The resulting Mel-scaled spectrograms have 64 frequency bins and 149 temporal units. Mel-scaled spectrograms are normalized to lie in the range of 0 to 1.

The missing samples of spectrograms are simulated by randomly masking 300 to 1500 ms of spectrograms’ temporal units. The total duration of masks is drawn from a normal distribution with a mean of 900 ms and a standard deviation of 300 ms. The masked area is split uniformly at random into 1 to 8 gaps of a minimum length of 36 ms. To recover the phase information of processed spectrograms and convert Mel spectrograms back into audio, the Griffin-Lim method is run for 300 iterations \cite{griffin}.

During video preparation, the RGB frames are converted into gray-scale images. Then, the dlib face detector is applied to get the frontal faces of the speakers and the dlib shape predictor is used to extract 68 facial landmarks \cite{dlib}. Using facial landmarks, we only keep mouth regions and crop video frames into areas of 100 × 50 pixels with the mouth in the center. Differences between facial landmarks of consecutive cropped frames are computed as visual features. This enables us to track motions between the frames, while throwing away the static background information. Finally, all processed visual features are normalized to have their values in the range of 0 to 1. 

To split the Grid Corpus into the train, validation, and test sets, 26 speakers (s1-20, s22-29, s31) are assigned to the train set, and 4 speakers (s30, s32-34) to the test. Moreover, the speakers (s26-27, s29, s31) are randomly split into two equal sets, with half of them being used as validation. The train and test splits are the same as the one used in \cite{Morrone} to provide us with a fair condition for comparisons. Hence, using 26 speakers for training and 4 (different) speakers for test will make 27752 and 1498 samples for train and test, respectively. Also, the validation set has 996 samples from a separate set of 4 speakers.

\subsection{Experiment Setting}

Throughout our training experiments in this section, we employ the Adam stochastic optimization algorithm \cite{adam} with a learning rate of 0.001 and a mini-batch size of 32 samples. During training, the learning rate is dropped by a factor of 0.1 in case the training loss stops improving for five epochs. Furthermore, an early stopping callback is defined that will terminate the training process if the validation loss does not decrease across ten consecutive epochs. The BLSTM layers have a latent dimension of 256 during all experiments in this paper. Our implementations are based on the Tensorflow Keras 2.11.0 and we run them on a MacBook Pro. M1 machine.

\subsection{Comparisons}

The core idea of our approach is to train an encoder-decoder model with BLSTM layers, which leverages pairs of visual and spectral features. Our model has been evaluated quantitatively on the Grid Corpus dataset. Similar to the method proposed in \cite{Morrone}, the reference and degraded audio quality and intelligibility are analyzed sample-by-sample via PESQ (Perceptual Evaluation of Speech Quality) \cite{pesq} and STOI (Short Term Objective Intelligibility) metrics \cite{stoi}. Also, the Peak Signal-to-Noise Ratio (PSNR) and the Mean Squared Error (MSE) are calculated to quantify the similarity of reconstructed spectrograms to the originals.

To investigate the effectiveness of the proposed approach, we compare it with several existing speech in-painting methods as listed in the below.
\begin{itemize}
  \item \textbf{A-SI}: The Audio Speech In-painting model is our baseline model for speech in-painting. It restores the masked spectrograms using only the input audio spectrograms. Similar to the decoder module of our sequence-to-sequence model depicted in Figure~\ref{fig:diag}, this model consists of a stack of three BLSTM layers and an FC layer. The FC is a time-distributed dense layer with 64 dimensions. The loss function of this model is MSE in order to minimize the error between generated and original spectrograms.
  \item \textbf{AV-S2S}: The Audio-Visual Sequence-to-Sequence model has the same architecture as our proposed model depicted in Figure~\ref{fig:diag}. It consists of a stack of three BLSTM layers in both the encoder and the decoder. We up-sample facial landmarks obtained from video frames to the same temporal length as audio spectrograms by applying a bivariate spline approximation. Then, the differences of consecutive facial landmarks of mouth regions are fed to BLSTM layers. The decoder takes the concatenation of the encoder's outputs and audio spectrograms as its inputs. The in-painted spectrograms are obtained via three BLSTM layers and a final FC layer. Unlike, the \textbf{AV-MTL-S2S} that is introduced in the below, here we only minimize the MSE loss function.
  \item \textbf{AV-MTL-S2S}: The Audio-visual Multi-Task Learning Sequence-to-Sequence model is the core model in this paper as illustrated in Figure~\ref{fig:diag}. Here, the loss function is as depicted in Equation~\ref{eq:loss} and the trade-off parameter $\lambda$ of the loss function is set to $0.001$. The activation function of all the FC layers is Relu.
   \item \textbf{AV-SI \cite{Morrone}}: We reproduce the Audio-Visual Speech In-painting model in \cite{Morrone} as a multi-modal speech in-painting baseline. Note that in the original paper, the visual features were the motion vectors of all facial landmarks, whereas we compute these features only on the mouth regions of video frames. The audio and visual features are temporarily synchronized, concatenated, and then fed into three layers of BLSTMs with the MSE loss. It is worth mentioning that the authors of \cite{Morrone} employed the $L_1$ loss function. 

\end{itemize}

Table~\ref{tab:comparison1} displays numerical results in terms of the mean values of the STOI, PESQ, PSNR, and MSE metrics over the test set. The first row of the table shows the average quality and intelligibility of the degraded input audio signals in comparison with the ground truth. The \textbf{A-SI} model has the lowest performance among studied approaches but still shows higher STOI and PESQ values when compared with the unprocessed input, in the first row. For the introduced corruptions of range 300 to 1500 ms, the \textbf{AV-MTL-S2S} model achieves the highest PESQ and STOI values, which indicates the highest quality and intelligibility of the restored audio signals. The significant improvement which have been gained in all metrics via implementing Audio-Visual models, i.e. \textbf{AV-S2S} and \textbf{AV-MTL-S2S}, demonstrates the beneficial role of visual modality for speech in-painting compared with the Audio-only one, i.e. \textbf{A-SI}.

It can be observed that the \textbf{AV-S2S} model performs similar to the \textbf{AV-SI} model of \cite{Morrone}, but, the results of \textbf{AV-MTL-S2S} surpass those from the \textbf{AV-SI \cite{Morrone}}. We believe this is due to the fact that we leverage a sequence-to-sequence model, which learns the sequence of visual features that are helpful for reconstructing spectrograms and Multi-Task Learning. Comparing \textbf{AV-MTL-S2S} with the \textbf{AV-S2S}, the improvement attained from Multi-Task Learning has a small margin with the Audio-Visual model. Here, all our metrics are computed on the entire signals or spectrograms, therefore we believe metrics such as STOI and PESQ show less sensitivity on average since most parts of the input signals are intact for small gaps of range less than 500 ms.  

The fact that our \textbf{AV-MTL-S2S} performs better than the \textbf{AV-S2S} model proves the usefulness of our multi-task learning approach in this problem. In other words, learnings from the phoneme recognition task can help with the speech in-painting task as well. 

The higher quality and intelligibility comes with the cost of higher training time and number of model parameters. While the training time of \cite{Morrone} is around two hours, both \textbf{AV-S2S} and \textbf{AV-MTL-S2S} need around four hours to be trained. The same trend is true for the number of trainable parameters with \cite{Morrone} having almost four million parameters and ours about nine million parameters.

\begin{figure}[t]
  \centering
  \includegraphics[width=\linewidth]{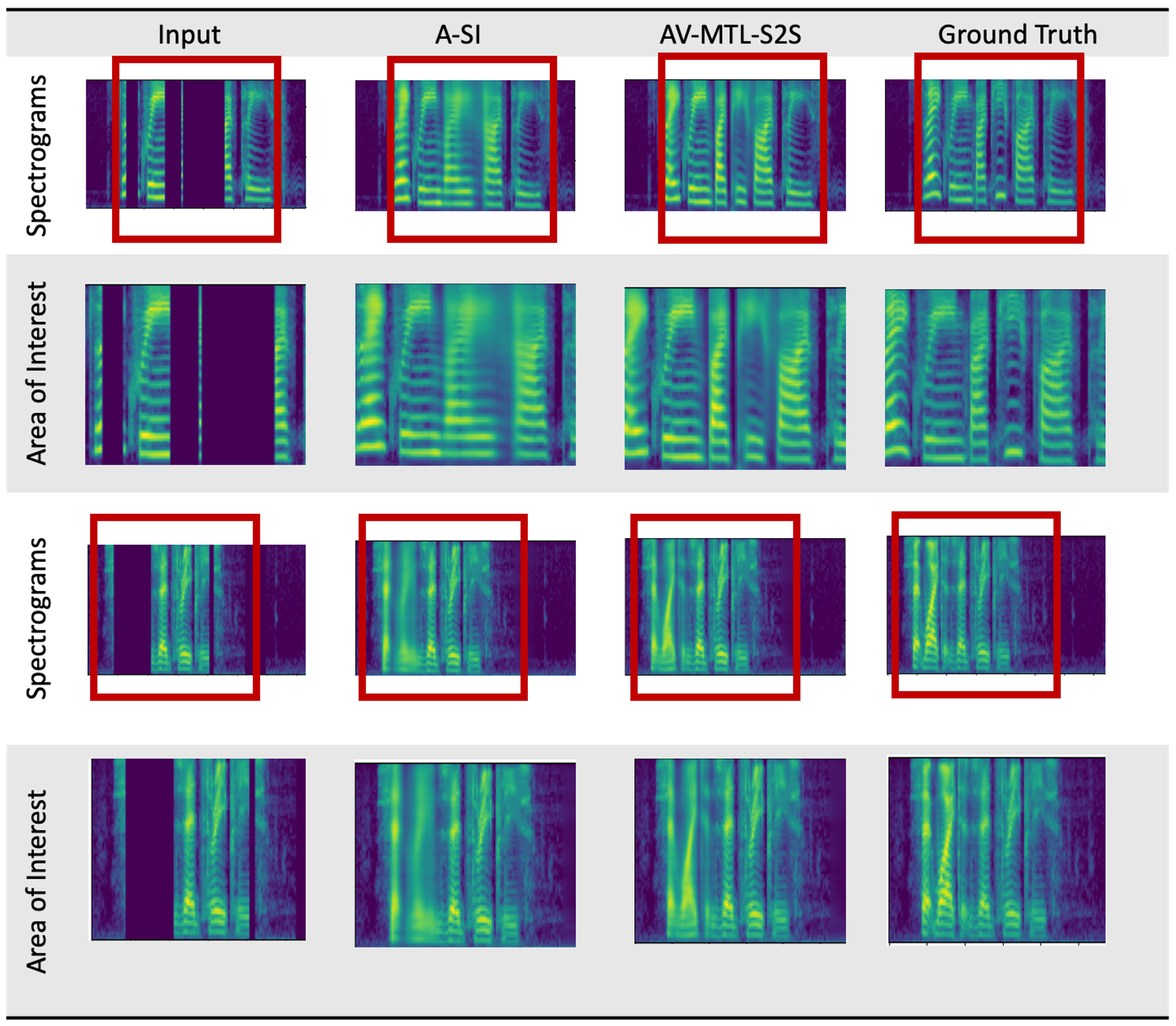}
  \caption{Qualitative results of in-painting distorted spectrograms for different methods. The masked areas are the areas of interest which are placed in red boxes and are zoomed-in for better visualization. The first two rows correspond to the first example and the last two are for the second example.}
  \label{fig:result}
\end{figure}

Figure~\ref{fig:result} shows the resulting spectrograms obtained from applying \textbf{A-SI} (column two) and \textbf{AV-MTL-S2S} (column three) to two example spectrograms. The input and ground-truth spectrograms are shown in columns one and four, respectively. The images suggest that the results from the multi-modal model surpass those of the audio-only model. We highlight the masked areas of interest in spectrograms by drawing red boxes. Zoomed-in views of the masked areas and their reconstructions are shown on the following lines. Comparing the images indicates that while the \textbf{A-SI} model fails to generate smooth spectrograms, our proposed \textbf{AV-MTL-S2S} generates visually reasonable and smooth results with more textual details. 

\begin{table}[t]
  \caption{A comparison of the speech in-painting methods studied in this paper in terms of STOI, PESQ, PSNR, and MSE. Upward arrows indicate higher values are better, while lower values are better for downward arrows.}
  \label{tab:comparison1}
  \centering
   \begin{tabular}{l | l l l l}
    \toprule
    \multicolumn{1}{c}{\textbf{}} & 
                                         \multicolumn{1}{c}{\textbf{PESQ}$^{\uparrow}$} & 
                                         \multicolumn{1}{c}{\textbf{STOI}$^{\uparrow}$} & 
                                         \multicolumn{1}{c}{\textbf{PSNR}$^{\uparrow}$}& 
                                         \multicolumn{1}{c}{\textbf{MSE}$^{\downarrow}$}\\
    \midrule
     Input                       & 1.60 & 0.63 & 14.07 & 0.046~~~\\
     A-SI             & 2.23  & 0.77 & 24.86 & 0.005~~~\\
     AV-S2S         & 2.31  &  0.80 & 25.58 & 0.004~~~ \\
     AV-MTL-S2S        & \textbf{2.33}& \textbf{0.80} & \textbf{25.80} & \textbf{0.004}~~~\\
     AV-SI \cite{Morrone}        & 2.30  & 0.79 & 25.52 & 0.004~~~\\

    \bottomrule
  \end{tabular}
\end{table}

\section{Conclusions}
\label{sec:con}

In this paper, we studied a novel multi-modal speech in-painter, which was inspired by the idea of sequence-to-sequence models in machine translation and image in-painting. The main contribution was to exploit the lip-reading and spectral features in an encoder-decode architecture. The visual cues were leveraged to guide speech generation. Our developed methods were evaluated on a speech dataset to regenerate the missing short words or phonemes. The duration of simulated distortions in our experiments range from 300ms to 1500ms for every 3s long audio. Distortions longer than 500ms are large and extremely large distortions (up to 1500ms) are studied in very few works. The experimental results showed the model was capable of restoring degraded spectrograms. Compared with the corrupted speech, both the quality and intelligibility of reconstructed audio pieces were improved. As a future direction, we plan to substitute RNN networks with transformer-based and covolutional models to do speech in-painting. Also, integrating a cross attention mechanism (similar to the one proposed in \cite{attention}) may help the decoder with attending to relevant visual frames for spectrogram in-painting. Furthermore, it is interesting to investigate how the results will be impacted if different noises are added to the audio, video frames are not clean and/or locations of missing audio parts are unknown.

\bibliographystyle{IEEEtran}
\bibliography{mybib}

\end{document}